\documentclass[preprint,showpacs,preprintnumbers,amsmath,amssymb]{revtex4}
\usepackage{graphicx}
\usepackage{dcolumn}
\usepackage{bm}
\usepackage{subfigure}
\begin{document}
\title{  FAST HADRON FREEZE-OUT GENERATOR, PART II: NONCENTRAL COLLISIONS.}
\author{N.S. Amelin}
\altaffiliation[deceased]{ }
\author{R. Lednicky}
\altaffiliation[Also at ]{
Institute of Physics AS CR, 18221 Praha 8, Czech Republic
}
\affiliation{Joint Institute for Nuclear Research, Dubna, Moscow Region,
141980, Russia}
\author{ I.P. Lokhtin}
\author{L.V. Malinina}
\altaffiliation[Also at ]{Joint Institute for Nuclear Research, Dubna, Moscow
Region, 141980, Russia}
\author{A.M. Snigirev}
\affiliation{M.V. Lomonosov Moscow State University, D.V. Skobeltsyn
Institute of Nuclear Physics, 119991, Moscow, Russia }
\author{Iu.A.~Karpenko}
\author{Yu.M. Sinyukov}
\affiliation{Bogolyubov Institute for Theoretical Physics, Kiev, 03143,
Ukraine}
\author{I.~Arsene}
\altaffiliation[Also at ]{
Institute for Space Sciences, Bucharest-Magurele, Romania
}
\author{L.~Bravina}
\affiliation{The department of Physics, University of Oslo, Norway}
\date{\today}
\begin{abstract}
The fast Monte Carlo procedure of  hadron
generation developed in our previous work is extended to describe noncentral
collisions of nuclei. We consider different possibilities to introduce
appropriate asymmetry
of the freeze-out hyper-surface and flow velocity profile.
For comparison with other models and experimental data
we demonstrate the results based on the standard parametrizations of the
hadron freeze-out hyper-surface and flow velocity profile
assuming either
a common chemical and thermal freeze-out or
the chemically frozen evolution from chemical to thermal
freeze-out.
The C++ generator code is written under the ROOT framework and is available
for public use at http://uhkm.jinr.ru/.
\end{abstract}
\pacs{25.75.Dw, 24.10.Lx, 25.75.Gz}

\maketitle
\section{\label{sec0}Introduction}
In the preceding work~\cite{FASTMC}
we have developed a Monte Carlo (MC) simulation procedure, and the corresponding
C++ code allowing for a fast but
realistic description of multiple hadron production in central relativistic
heavy ion collisions.
A high generation speed and  easy control through input
parameters make our MC generator code particularly useful
for detector studies.
The generator code is quite flexible and allows the user to add
other scenarios and freeze-out surface parametrizations as well as
additional hadron species in a simple manner.
We have compared the  BNL Relativistic Heavy
Ion Collider (RHIC) experimental data on central Au+Au collisions
with our MC
generation results obtained within the single freeze-out scenario with
Bjorken-like and Hubble-like freeze-out surface parametrizations.
Although simplified, such a scenario nevertheless
allowed for a reasonable description of particle spectra
and femtoscopic momentum correlations.
This description can be farther improved by introducing finite
emission duration and extending the table of the included
resonances;
the single freeze-out scenario is however less successful in the
description of the data on elliptic flow (see section III).

The particle densities at the chemical freeze-out
stage are too high (see, e.g.,~\cite{Sinyukov02})
to consider particles as free streaming and to associate this stage
with the thermal freeze-out one.
In this work we have implemented as an option
more sophisticated scenario of thermal freeze-out: the system expands hydrodynamically with
frozen chemical composition, cools down and finally decays at some thermal
freeze-out hypersurface. The RHIC experimental data are compared
with our MC generation results obtained within this
thermal freeze-out scenario.
We do not consider here a more complex freeze-out scenario
taking into account continuous particle emission (see, e.g.,~\cite{Borysova}).

In present paper, we also extend the fast Monte Carlo procedure of  hadron
generation developed in our previous work~\cite{FASTMC}
to describe noncentral collisions of nuclei.
One of the most
spectacular features of the RHIC data is large
elliptic flow~\cite{v2_STAR}. The development of a strong flow is well described by the hydrodynamic
models and requires short time scale and large pressure gradients,
attributed to strongly interacting systems.
However, results of
hydrodynamic models significantly disagree with the
data on femtoscopic momentum correlations (compare~\cite{CF_STAR}
with, e.g.,~\cite{Kolb_Heinz}), related
with the space-time characteristics of the system at freeze-out.
Usually, the hadronic cascade models underestimate the momentum anisotropy
and overestimate the source sizes (e.g.~\cite{Voloshin-Hardtke,Voloshin-flow,Bleicher02}). Some
sophisticated hybrid models (e.g. AMPT~\cite{AMPT})
reproduce the elliptic flow and the correlation radii
but with different sets of model parameters.

Successful attempts to describe simultaneously the momentum-space measurements and
the freeze-out coordinate-space data were done in several models which
make experimental data fitting within some parametrizations of freeze-out hypersurface:
``Kiev-Nantes'' model~\cite{Borysova}, ``Blast-Wave''
parametrizations~\cite{Lisa, THERMINATOR, THERMINATOR2},
 ``Buda-Lund'' hydro approach~\cite{Buda-Lund}.
All these approaches use the
hydro-inspired parametrizations of freeze-out hypersurface and help in understanding
the full freeze-out scenario at RHIC.

In this article we analyze the RHIC data at $\sqrt{s}_{NN}=200$ GeV and
try to use the same set of the model parameters
for the description of both the momentum-space observables, i.e.
transverse mass ($m_t$) spectra and $p_t$-dependence of elliptic flow, and
freeze-out coordinate-space observables, i.e. $k_t$-dependence and
azimuthal angle ($\Phi$) dependence of the correlation radii.
The chemical composition of the fireball was fixed in our previous article~\cite{FASTMC} by
the particle ratios analysis.

The paper is organized as follows. Section~\ref{sec1}  is devoted to the
description of main modifications of the model~\cite{FASTMC} needed to take
into consideration noncentral collisions.
In section~\ref{sec2}
the example calculations are compared  with the
RHIC experimental data. We summarize and conclude in
section~\ref{sec3}.

\maketitle
\section{\label{sec1}Freeze-out surface parametrizations}

The extension of our MC generator to noncentral collisions demands mainly the
modifications of freeze-out hypersurface parametrizations (Sec. V of
Ref.~\cite{FASTMC}) and does not practically influence the generation
procedure itself (Sec. VI of Ref.~\cite{FASTMC}). Therefore we focus on these
modifications only considering the popular
Bjorken-like and Hubble-like freeze-out parametrizations respectively used
in so-called  blast wave~\cite{Lisa} and Cracow~\cite{Bron04} models
as the example options in our MC generator.
Similar parametrizations have been
used in the hadron generator THERMINATOR
\cite{THERMINATOR}.

As usual, in the Bjorken-like parametrization, we
substitute the Cartesian coordinates $t$, $z$ by
the Bjorken ones~\cite{Bjorken83}
\begin{equation}
\label{FS2}
\tau = (t^2 - z^2)^{1/2},~~
\eta = \frac{1}{2} \ln \frac{t+z}{t-z},
\end{equation}
and introduce the the radial
vector $\vec{r}\equiv\{x,y\}=\{r \cos \phi, r \sin \phi\}$, i.e.,
\begin{equation}
\label{xB}
x^{\mu} =
\{\tau \cosh \eta , \vec{r} , \tau \sinh\eta\}=
\{\tau \cosh \eta ,r \cos \phi, r \sin \phi , \tau \sinh\eta\}.
\end{equation}

For a freeze-out hypersurface represented
by the equation $\tau=\tau(\eta,r,\phi),$ the
hypersurface element in terms of the coordinates $\eta$, $r$,
$\phi$ becomes
\begin{equation}
\label{FS1}
d^3\sigma_{\mu}= \epsilon_{\mu
\alpha \beta \gamma} \frac{dx^{\alpha}dx^{\beta}dx^{\gamma}} {d\eta dr
d\phi}d\eta dr d\phi,
\end{equation}
where $\epsilon_{\mu \alpha \beta\gamma}$ is the completely
antisymmetric Levy-Civita tensor in four dimensions with
$\epsilon^{0123} = -\epsilon_{0123} = 1$.
Generally, the freeze-out hypersurface is represented by a set of equations
$\tau=\tau_{j}(\eta, r, \phi)$ and Eq.~($\ref{FS1}$) should be substituted by
the sum of the corresponding hypersurface elements.
For the simplest and frequently used freeze-out hypersurface $\tau = const$,
one has
\begin{equation}
\label{FS5}
\begin{array}{c}
d^3\sigma_{\mu}=n_{\mu}d^3\sigma=\tau d^2\vec{r}d\eta
\{\cosh\eta,0,0,
-\sinh\eta \},
\cr
d^3\sigma=\tau d^2\vec{r}d\eta,
\cr
n^{\mu} =\{
\cosh\eta,0,0,
\sinh\eta \}.
\end{array}
\end{equation}

In noncentral collisions the shape of the emission region in
the transverse ($x$-$y$) plane
can be approximated by an ellipse
(as usual, the $z$-$x$ plane coincides with the reaction plane).
The ellipse radii $R_x(b)$ and $R_y(b)$ at a given impact
parameter $b$ are
usually parametrized~\cite{Lisa, huovinen01, wiedemann98,
broniowski02} in terms of the spatial anisotropy
$\epsilon(b)=(R_y^2-R_x^2)/(R_x^2+R_y^2)$ and the
scale factor $R_s(b)=[(R_x^2+R_y^2)/2]^{1/2}$,
\begin{equation}
\label{Rxy}
R_x(b) = R_s(b) \sqrt{1-\epsilon(b)},~~~ R_y(b) = R_s(b) \sqrt{1+\epsilon(b)}
.
\end{equation}
Then from the ellipse equation,
\begin{equation}
\label{ellipse}
\frac{x^2}{R^2_x} + \frac{y^2}{R^2_y} = 1,
\end{equation}
follows the explicit dependence of
the fireball transverse radius $R(b,\phi)$ on
the azimuthal angle
$\phi$:
\begin{equation}
\label{Rbphi}
R(b,\phi)= R_s(b)
 \frac{\sqrt{1-\epsilon^2(b)}} {\sqrt{1+\epsilon(b) \cos2\phi}}~;
\end{equation}
particularly, $R(b,0)=R_x(b)$ and $R(b,\pi/2)=R_y(b)$.
To reduce the number of free parameters, we assume here
a simple scaling option~\cite{Lokhtin05}
\begin{equation}
\label{Rb}
R_s(b) = R_s(b=0) \sqrt{1-\epsilon_{s}(b)},
\end{equation}
where $R_s(b=0)\equiv R$ is the fireball freeze-out transverse radius in central collisions.
It means that the
dimensionless ratio $R_s(b)/R_s(0)$ at the
freeze-out moment
depends on the collision energy, the radius $R_A$ of the
colliding (identical) nuclei and the impact
parameter $b$ through a dimensionless $\epsilon_{s}(b)$
only. It should be noted that both
$\epsilon_{s}(b)$ and the fireball freeze-out eccentricity $\epsilon(b)$
are determined by the eccentricity
$\epsilon_{0}(b)=b/(2R_A)$
of the elliptical overlap of the colliding nuclei
at the initial moment, when
\begin{equation}
\label{scaling}
\frac{R_s(b)}{R_s(b=0)}\Big|_{\epsilon(b)=\epsilon_0(b)}
\equiv  \frac{R_s(b)_{\rm initial}}
{R_A} = \sqrt{1-\epsilon_{0}(b)} .
\end{equation}
Since $\epsilon_{s}(0)=\epsilon(0)=\epsilon_{0}(0)=0$,
one can can assume that $\epsilon_{s}(b) \simeq \epsilon(b)$
at sufficiently small values of the impact parameter $b$.
It appears that the use of the simple ansatz $\epsilon_{s}(b)=\epsilon(b)$
allows one to achieve the absolute normalization of particle spectra correct
within $\sim 10 \%$ up to $b \simeq R_{A}$ (see section~\ref{sec23}).

If the system evolution were driven by the pressure gradients, the
expansion would be stronger in the direction of the short ellipse
$x$-axis (in the reaction plane), where the pressure gradient is
larger than in the direction of the long ellipse $y$-axis (see,
e.g.,~\cite{Kolb_Heinz}). The typical hydrodynamic evolution
scenario is shown in Fig.~$\ref{fig:hydro_evolution}$. During the
evolution, the initial system coordinate anisotropy
$\epsilon_0(b)$ is transformed into the momentum anisotropy $ \delta(b) $.
According to the hydrodynamical calculations, the spatial
eccentricity almost disappears and the momentum anisotropy
saturates at rather early evolution stage before freeze-out.
As we do not trace the evolution here,
we will consider the spatial and momentum anisotropies
$\epsilon(b)$ and $ \delta(b) $  as free parameters.

For central collisions the fluid flow four-velocity
$u^{\mu}(t, \vec{x})= \gamma(t, \vec{x}) \{1,\vec{v}(t, \vec{x})\}
\equiv \gamma(t, \vec{x}) \{1,\vec{v}_{\perp}(t, \vec{x}),v_z(t, \vec{x})\}$
at a point $\vec{x}$ and time $t$ was parametrized~\cite{FASTMC}
in terms of the
longitudinal ($z$) and transverse ($\perp$) fluid flow rapidities
\begin{equation}
\label{HC2}\eta_u(t, \vec{x}) = \frac{1}{2}
\ln{\frac{1+v_z(t, \vec{x})}{1-v_z(t, \vec{x})}},~~
\rho_u(t, \vec{x}) = \frac{1}{2}
\ln{\frac{1+v_{\perp}(t, \vec{x})\cosh\eta_u(t, \vec{x})}{1-v_{\perp}(t, \vec{x})\cosh\eta_u(t, \vec{x})}},
\end{equation}
where
$v_{\perp} = | \vec{v}_{\perp}|$ is the magnitude of the transverse component of the flow
three-velocity
$ \vec{v}=
\{ v_{\perp} \cos \phi_u , v_{\perp} \sin \phi_u, v_z\}$,
i.e.,
\begin{equation}
\begin{array}{c}
\label{HC6}u^{\mu}(t, \vec{x})=
\{\cosh \rho_u \cosh \eta_u, \sinh \rho_u \cos \phi_u, \sinh \rho_u \sin \phi_u,
\cosh \rho_u \sinh \eta_u \}
\\
=\{
 (1+u_{\perp}^{2})^{1/2} \cosh \eta_u,\vec{u}_{\perp}, (1+u_{\perp}^{2})^{1/2} \sinh \eta_u\},
\end{array}
\end{equation}
$\vec{u}_{\perp} = \gamma\vec{v}_{\perp}= \gamma_{\perp}\cosh\eta_u \vec{v}_{\perp}$,
$\gamma_{\perp}=\cosh \rho_u$.
However, unlike the transverse isotropic parametrization ($\phi_u=\phi$), now
the azimuthal angle $\phi_u$ of the fluid velocity vector is not necessarily
identical to the spatial azimuthal angle $\phi$, because of the
nonzero flow anisotropy parameter $\delta(b)$ ~\cite{wiedemann98,broniowski02}
:
\begin{equation}
\begin{array}{c}
\label{HC6b}u^{\mu}(t,\vec{x})=
\{\gamma_{\phi}\cosh \tilde \rho_u \cosh \eta_u, \sqrt{1+\delta(b)}\sinh \tilde \rho_u \cos
\phi, \\
\sqrt{1-\delta(b)}\sinh \tilde \rho_u \sin \phi,
\gamma_{\phi}\cosh \tilde \rho_u \sinh \eta_u \},
\end{array}
\end{equation}
where
\begin{equation}
\label{gammaphi}
\gamma_{\phi}=
\sqrt{1+\delta(b) \tanh^2 \tilde \rho_u \cos 2\phi},
\end{equation}
\begin{equation}
\label{tanphi}
\tan\phi_u=
\sqrt{\frac{1-\delta(b)}{1+\delta(b)}}\tan\phi.
\end{equation}
The transverse flow rapidity $\rho_u$ is related to $\tilde \rho_u$
by:
\begin{equation}
\label{u_perp}
u_{\perp}= \sinh \rho_u=
\sqrt{1+\delta(b) \cos 2\phi}~ \sinh \tilde \rho_u.
\end{equation}
Note, that for $\delta(b)=0$ (i.e. $\phi_u=\phi$), Eq. (\ref{HC6b})
reduces to Eq. (\ref{HC6}) which
was applied in Refs.~\cite{Lokhtin05, heinz02}.
In Ref.~\cite{broniowski02}, $\delta(b)$ is obtained by fitting the model prediction
to the measured elliptic flow coefficient $v_2$.

Further we assume the longitudinal boost invariance~\cite{Bjorken83}
$\eta_u = \eta$, which is a good approximation for the highest RHIC energies at
the midrapidity region.
To account for the violation of the boost invariance, we have also
included in the code an option corresponding to the substitution
of the uniform distribution of the space-time longitudinal
rapidity $\eta$ in the interval $[-\eta_{\max},\eta_{\max}]$
by a Gaussian distribution $\exp(-\eta^2/2\Delta\eta^2)$
with a width parameter $\Delta\eta=\eta_{max}$.
The presence of the ``oscillation term''
$ \sqrt{1+\delta(b) \cos 2\phi}$ in the transverse component $u_{\perp}$
of the flow velocity in Eq.~(\ref{u_perp})
allows us to use the simple linear profile
for $\tilde \rho_u$ without introduction of the additional parameters for each
centrality ($b$) unlike other models, namely:
 \begin{equation}
\label{URBaN}
\tilde \rho_u=\frac{r}{R_{s}(b)}\rho_u^{\rm max}(b=0),
\end{equation}
where $\rho_u^{\rm max}(b=0)$ is the maximal transverse flow rapidity for central
collisions. At such normalization and $\delta(b)>\epsilon(b)$ the
maximal transverse flow ($u_{\perp}, \rho_u$)
is achieved at $\phi=0$, i.e.
along $x$-axix as it
should be according to the
hydrodynamic scenario described above (Fig.~\ref{fig:hydro_evolution}).
(although $\tilde \rho_u$ has a maximum at  $\phi=\pi/2$!)

Here one should note that
the ``popular parametrization'' of transverse flow rapidity used in Ref.~\cite{Lisa}
(and implemented as an option in our MC generator also):
\begin{equation}
\label{rhooscillation}
\rho_u = \tilde{r}[\rho_0(b)+\rho_2(b)\cos 2\phi_u],
\end{equation}
where
\begin{equation}
\label{roscillation}
\tilde{r} \equiv \sqrt{\Bigl(\frac{r\cos\phi}{R_x}\Bigr)^2 +
\Bigl(\frac{r\sin\phi}{R_y}\Bigr)^2} =\frac{r}{R(b,\phi)}
\end{equation}
is the ``normalized elliptical radius'', $\rho_0(b)$ and $\rho_2(b)$ are the two
fitting parameters,
is close to our  parametrization and gives the similar results for observables
under consideration.
In  parametrization of Ref.~\cite{Lisa}
the boost is perpendicular to the
elliptical subshell on which the source element is found:
$\tan \phi_u = (R_{x}^{2}/R_{y}^{2}) \tan \phi =
(1-\epsilon)/(1+\epsilon) \tan \phi$
and $\delta(b)=2\epsilon(b)/(1+\epsilon^2(b))$.
It is interesting to note that for sufficiently weak transverse
flows, $\rho_u \leq 1$, considered here, one can put
$\sinh\rho_u \simeq \rho_u$ and obtain our parametrization
from that of Ref.~\cite{Lisa} by substitutions
\begin{equation}
\label{connection}
\frac{\rho_0(b)}{R(b,\phi)}\rightarrow \frac{\rho_u^{\max}(b=0)}{R_{s}(b)}~~
~~~
1+ \frac{\rho_2(b)}{\rho_0(b)} \cos 2\phi_u \rightarrow \sqrt{1+\delta(b) \cos
2\phi}.
\end{equation}
Thus, in the case of moderate transverse flows,
one can obtain the same result either by fixing the direction
of the flow velocity vector but allowing for the azimuthal dependence
of the flow rapidity or by allowing for arbitrary direction
of the flow velocity vector but assuming
azimuthally independent flow rapidity.

At $\tau = const$,
the total effective volume for particle
production in the case of noncentral collisions becomes
\begin{eqnarray}
\label{HD3} V_{\rm eff}=\int_{\sigma(t, \vec{x})} d^3\sigma_{\mu}(t, \vec{x})u^{\mu}(t, \vec{x}) =
\tau \int_0^{2\pi} d\phi \int_0^{R(b,\phi)} (n_{\mu}u^{\mu}) r dr
\int_{\eta_{\min}}^{\eta_{\max}} d\eta,
\end{eqnarray}
where $(n_{\mu}u^{\mu})=\cosh \tilde \rho_u
\sqrt{1+\delta(b) \tanh^2 \tilde \rho_u \cos 2 \phi}$ .

We also consider the Cracow model scenario \cite{Bron04}
corresponding to the Hubble-like freeze-out hypersurface
$\tau_H=(t^2-x^2-y^2-z^2)^{1/2}=const$.
Introducing the longitudinal space-time rapidity $\eta$ according to Eq.~(\ref{FS2})
and the transverse space-time rapidity $\rho=\sinh^{-1}(r/\tau_H)$,
one has \cite{Csorgo96}
\begin{equation}
\label{xH}
x^{\mu}=\tau_H\{\cosh \eta \cosh \rho, \sinh \rho \cos \phi, \sinh \rho
\sin \phi, \sinh \eta \cosh \rho \},
\end{equation}
$\tau_H=\tau_B/\cosh \rho$.
Representing the freeze-out hypersurface by the equation
$\tau_H=\tau_H(\eta,\rho,\phi)=const$, one finds from
Eq.~($\ref{FS1}$):
\begin{equation}
\begin{array}{{c}}
\label{cracow}
d^3\sigma=\tau^{3}_H\sinh \rho \cosh \rho d \eta d \rho d \phi =\tau_H
d \eta d^2\vec{r},
\\
n^{\mu}(t, \vec{x})=x^{\mu}(t, \vec{x})/\tau_H.
\end{array}
\end{equation}
With the additional flow anisotropy parameter $\delta(b)$ the flow four-velocity
is parametrized as~\cite{broniowski02}:
\begin{equation}
\begin{array}{c}
\label{huble}u^{\mu}(t, \vec{x})=
\{\gamma_{\phi}^H\cosh \rho\cosh \eta, \sqrt{1+\delta(b)}\sinh \rho \cos
\phi, \\
\sqrt{1-\delta(b)}\sinh \rho \sin \phi,
\gamma_{\phi}^H\cosh \rho \sinh \eta \},
\end{array}
\end{equation}
where
\begin{equation}
\label{gammaphiH}
\gamma_{\phi}^H=
\sqrt{1+\delta(b) \tanh^2 \rho \cos 2 \phi}.
\end{equation}
The effective volume corresponding to $r=\tau_H \sinh \rho<R(b,\phi)$
and $\eta_{\min} \le \eta \le \eta_{\max}$
is
\begin{equation}
\label{HD5}
V_{\rm eff}=\int_{\sigma(t, \vec{x})}d^3 \sigma_{\mu}(t, \vec{x})u^{\mu}(t, \vec{x}) = \tau_H
\int_{0}^{2 \pi} d \phi \int^{R(b,\phi)}_{0} (n_{\mu}u^{\mu})
r dr  \int_{\eta_{\min}}^{\eta_{\max}}d \eta
\end{equation}
with
\begin{equation}
\begin{array}{c}
(n_{\mu}u^{\mu})=  \cosh^ 2\rho \Bigl(
\sqrt{1+\delta(b) \tanh^2 \rho \cos 2 \phi }\\
-\tanh^2 \rho (\sqrt{1+\delta(b)} \cos^2 \phi +
\sqrt{1-\delta(b)} \sin^2 \phi)\Bigr) \simeq 1+o(\delta^2(b)).
\end{array}
\end{equation}

Our MC procedure to generate the freeze-out
hadron multiplicities, four-momenta and four-coordinates
for central collisions
has been described in detail in Ref.~\cite{FASTMC}.
For noncentral collisions, only the
generation of the transverse radius $r$ is slightly different,
taking
place in the azimuthally dependent interval $[0, R(b, \phi)]$.

\maketitle
\section{\label{sec2}Input parameters and example calculations}

\subsection{\label{sec21} Model input parameters}
First, we summarize the input parameters which control the
execution of our MC hadron generator
in the case of Bjorken-like and Hubble-like parametrizations, and
should be specified for different energies, ion beams and event centralities.
\begin{enumerate}
\item  Thermodynamic parameters at chemical freeze-out: temperature
$T^{\rm ch}$ and chemical potentials per a unit charge
$\widetilde{\mu}_B, \widetilde{\mu}_S, \widetilde{\mu}_Q$.
As an option,  an additional parameter $\gamma_s \le 1$
takes into account the strangeness suppression according to the partially
equilibrated distribution \cite{Gorenstein97,Raf81}:
\begin{equation}
f_{i}(p^{*0};T,\mu_i,\gamma_s) = \frac{g_i}{\gamma_s^{-n_i^s}\exp{([p^{*0} -
\mu_i]/T)} \pm 1},
\end{equation}
where $n_i^s$ is the number of strange quarks and
antiquarks in a hadron $i$, $p^{*0}$ is the hadron energy in the fluid element
rest frame,
$g_i=2 J_i+1$
is the spin degeneracy factor
Optionally, the parameter $\gamma_s$ can be fixed using its
phenomenological dependence on the temperature and
baryon chemical potential \cite{Beca06}.
\item
Volume parameters:
the fireball transverse radius $R(b=0)$ (determined in central collisions;
in noncentral collisions we use the scaling option (\ref{Rb},\ref{scaling})
to recalculate $R(b)$ from $R(b=0)$),
the freeze-out proper time $\tau$ and its standard deviation $\Delta \tau$
(emission duration) \cite{Bravina}.
\item  Maximal transverse flow rapidity $\rho_u^{\max}(b=0)$
for Bjorken-like parametrization   in central collisions.
\item  Maximal space-time longitudinal rapidity $\eta_{\max}$ which
determines the rapidity interval $[-\eta_{\max}, \eta_{\max}]$
in the collision center-of-mass system.
To account for the violation of the boost invariance, we have
included in the code an option corresponding to the substitution
of the uniform distribution of the space-time longitudinal
rapidity $\eta$ in the interval $[-\eta_{\max},\eta_{\max}]$
by a Gaussian distribution $\exp(-\eta^2/2\Delta\eta^2)$
with a width parameter $\Delta\eta=\eta_{max}$
(see, e.g., \cite{Wiedemann_Heinz02, Lokhtin05}).
\item Impact parameter range: minimal $b_{\min}$ and maximal $b_{\max}$ impact parameters.
\item Flow anisotropy parameter $\delta(b)$
in Bjorken-like and Hubble-like parametrizations (or $\rho_0(b)$ and $\rho_2(b)$ in
the ``Blast-Wave'' parametrization of Ref.~\cite{Lisa}).
\item Coordinate anisotropy parameter $\epsilon(b)$.
\item Thermal freeze-out temperature $T^{\rm th}$ (if single freeze-out is considered, $T^{\rm th}=T^{\rm ch}$).
\item Effective chemical potential of $\pi^{+}$ at thermal freeze-out $\mu_{\pi}^{\rm eff~th}$ (0, if single freeze-out is considered).
\item Parameter which enables/disables weak decays.
\end{enumerate}

 \begin{table}
 \caption{\label{tab:table1}
Model parameters for  central Au + Au
collisions at $\sqrt s_{NN} = 200$~GeV. Chemical freeze-out parameters: $T^{\rm
ch}$=0.165~GeV,
$\widetilde{\mu}_{B}$=0.028~GeV, $\widetilde{\mu}_{S}$=0.007~GeV,
$\widetilde{\mu}_{Q}$= -- 0.001~GeV.
 }
 \begin{ruledtabular}
 \begin{tabular}{cccccc}
 $T^{\rm th}$, GeV  & 0.165 & 0.130 &  0.100 \\
 \colrule
$\tau$, fm/$c$         & 7.0     & 7.2 & 8.0     \\
$\Delta \tau $, fm/$c$ & 2.0     & 2.0  & 2.0    \\
$R(b=0)$,  fm        & 9.0     & 9.5 & 10.0    \\
$\rho_u^{\max}(b=0)$ & 0.65    & 0.9  & 1.1    \\
$\mu_{\pi}^{\rm eff~th}$ & 0.      & 0.10  & 0.11  \\

\end{tabular}
 \end{ruledtabular}
 \end{table}

The parameters used to simulate central collisions are given in Table \ref{tab:table1}.
The parameters determined in central collisions for  $T^{\rm th}$=0.1~GeV:
$\tau$=8.0 fm/$c$, $R(b=0)$=10.~fm, $\Delta \tau $=2.0~fm/$c$;
$\rho_u^{\max}(b=0)=1.1$ (3-th column in Table \ref{tab:table1}) were used to
simulate
 Au+Au collisions at $\sqrt s_{NN} = 200$~GeV at different
centralities. The additional parameters needed only for noncentral collisions
are given
in Table \ref{tab:table2}.

\begin{table}
 \caption{\label{tab:table2}
Model parameters for Au + Au
collisions at $\sqrt s_{NN} = 200$~GeV at different centralities.
Chemical freeze-out parameters: $T^{\rm ch}$=0.165~GeV,
$\widetilde{\mu}_{B}$=0.028~GeV, $\widetilde{\mu}_{S}$=0.007~GeV,
$\widetilde{\mu}_{Q}$= -- 0.001~GeV.
Thermal freeze-out parameters: $T^{\rm th}$=0.1~GeV, $\mu_{\pi}^{\rm eff~th}$=0.11~GeV.
Volume parameters determined in the central collisions: $R(b=0)=$10.0~fm, $\tau=$8.0~fm/$c$, 
$\rho_u^{\max}(b=0)=1.1$
 }
 \begin{ruledtabular}
 \begin{tabular}{ccccccc}
 centrality   & c=0--5 $\%$ & c=5--10 $\%$  & c=10--20 $\%$ & c=20--30 $\%$  &
 c=30--40 $\%$ & c=40--60 $\%$  \\
\colrule
$b_{\min}/R_A$      & 0.    & 0.447 & 0.632  & 0.894& 1.095& 1.265 \\
$b_{\max}/R_A$      & 0.447 & 0.632 & 0.894  & 1.095& 1.265& 1.549 \\
$\epsilon(b)$  &0     & 0    & 0     & 0.1  & 0.15 & 0.15\\
$\delta(b)$    & 0.05  & 0.08  & 0.12   & 0.25 & 0.34 & 0.36  \\

\end{tabular}
 \end{ruledtabular}
 \end{table}

\maketitle
\subsection{\label{sec22} Different chemical and thermal freeze-outs}
Since the assumption of a common chemical and thermal freeze-out
can hardly be justified (see, e.g.,~\cite{Sinyukov02}), we consider
here a more complicated
scenario with different chemical and thermal freeze-outs.

The mean particle numbers $\bar{N}_i^{\rm th}$ at thermal
freeze-out can be determined using the following procedure \cite{Sinyukov02}.
In our preceding article \cite{FASTMC} the temperature and chemical
potentials at chemical freeze-out have
been fixed by fitting the ratios of the numbers of (quasi)stable particles.
The common factor, $V_{\rm eff}^{\rm ch}$,
and, thus, the absolute particle and resonance numbers was fixed
by pion multiplicities.
Within the concept of chemically
frozen evolution these numbers are
assumed to be conserved
except for corrections due to decay of some
part of short-lived resonances that can be estimated from the assumed
chemical to thermal freeze-out evolution time.
Then one can calculate the mean numbers of different
particles and resonances reaching a (common) thermal freeze-out
hypersurface. At a given thermal freeze-out temperature $T^{\rm th}$ these mean
numbers can be expressed through the thermal effective volume $V_{\rm eff}^{\rm th}$
and the chemical
potentials for each particle species $\mu^{{\rm th}}_{i}$. The latter can no longer
be expressed in the form $\mu_i =\vec{q}_i\vec{\widetilde{\mu}}$, which is valid only for
chemically equilibrated systems. For a given parametrization of the thermal
freeze-out hypersurface, the thermal effective volume $V_{\rm eff}^{\rm th}$
(and thus all $\mu^{{\rm th}}_{i}$)
can be fixed with the help of pion interferometry data.

In practical calculations
the particle number density $\rho_{i}^{\rm eq}(T, \mu_i)$
is represented in the form of a fast converging series \cite{FASTMC}:
\begin{equation}
\label{GC4} \rho_{i}^{\rm eq}(T, \mu_i) = \frac{g_i}{2 \pi^2}m^2_iT\sum_{k=1}^{\infty}
\frac{(\mp)^{k+1}}{k} \exp(\frac{k\mu_i}{T})K_2(\frac{km_i}{T}),
\end{equation}
where $K_2$ is the modified Bessel function of the
second order, $m_i$ and $g_i=2 J_i+1$ are the mass and
the spin degeneracy factor of particle $i$ respectively.

Using Eq.~(\ref{GC4}) and the assumption of the conservation of
the particle number ratios from the
chemical to thermal freeze-out evolution time,
we obtain the following ratios for $i$-particle specie
to $\pi^{+}$:
\begin{equation}
 \label{GC5}\frac {\rho_{i}^{\rm eq}(T^{\rm ch}, \mu_i)}
 {\rho_{\pi}^{\rm eq}(T^{\rm ch},\mu_i^{\rm ch})}
= \frac{\rho_{i}^{\rm eq}(T^{\rm th}, \mu_i^{\rm th})}{\rho_{\pi}^{\rm eq}(T^{\rm th}, \mu_{\pi}^{\rm eff~th})}.
\end{equation}

The absolute values of particles densities $\rho_{i}^{\rm
eq}(T^{\rm th}, \mu_{i}^{\rm th})$ are determined by the choice of
the free parameter of the model: effective pion chemical potential
$\mu_{\pi}^{\rm eff~th}$ at the temperature of thermal freeze-out
$T^{\rm th}$.
Assuming for the other particles (heavier then pions) the
Boltzmann approximation in Eq.~(\ref{GC4}) one deduces from
Eqs.~(\ref{GC4}) - (\ref{GC5}) the chemical potentials of
particles and resonances at thermal freeze-out:
\begin{equation}
\label{GC7c} \mu_{i}^{\rm th} = T^{\rm th}\ln(\frac{\rho_{i}^{\rm
eq} ( T^{\rm ch}, \mu_{i}^{\rm ch})}{\rho_{i}^{\rm eq}(T^{\rm
th},\mu_i=0)}\frac{\rho_{\pi}^{\rm eq} ( T^{\rm th},
\mu_{\pi}^{\rm eff~th})}{\rho_{\pi}^{\rm eq}(T^{\rm ch},\mu_i^{\rm
ch})}).
\end{equation}

The correct way to determine the best set of model parameters
would be achieved by fitting all the observables together as it
was suggested in Ref.~\cite{Wiedemann_Heinz02}, but for our
MC-type model it is technically impossible. For the example
calculations with our model at RHIC energies we choose
$T^{\rm ch} = 0.165$~GeV and the thermal
temperatures as in the analytical models which performed the
successful fitting of RHIC data: $T^{\rm th}=T^{\rm ch}=0.165$~GeV
(Cracow model \cite{Bron04}) and $T^{\rm th}=0.100$~GeV
(Blast-Wave model \cite{Lisa}), and some arbitrary intermediate
temperature $T^{\rm th}=0.130$~GeV. It is well known (see,
e.g.,~\cite{Sinyukov02}) that the pion transverse spectra at
thermal freeze-out can be described in two regimes: low
temperature and large transverse flow on the one hand, and higher
temperature and non-relativistic transverse flow on the other hand
(see section~\ref{sec23}).
The low temperature regime seems to be preferable because
the strong transverse flow is expected to describe the
large inverse slopes of transverse spectra
of the heavy hadrons (especially protons) and small correlation radii
obtained at RHIC better \cite{Lisa, Borysova}.
We present the calculated correlation radii in section~\ref{sec25}.

In the considered here last version of FASTMC the new table of resonances was included.
It contains 360 resonances and stable particles, instead of 85 ones included
in the previous versions. This particle table is produced from the SHARE \cite{SHARE}
particle table excluding not well established resonances states.
The decays of resonances are controlled by its lifetime $1/\Gamma$, there
$\Gamma$ is the width of resonance specified in the particle table, and
 they
occur with the probability density $\Gamma exp(-\Gamma \tau)$ in the resonance rest frame.
Then the decay products are boosted
to the reference frame in which the freeze-out hypersurface was defined.
Because we need to compare our calculations with data from different experiments
we made possible to switch on/off different decays based on their lifetime (i.e. turn on/off weak decays).
Only the two- and three-body decays are considered in our model. The branching ratios
are also taken from the particle decay table produced from the SHARE decay table
\cite{SHARE}. The cascade decays are also possible.

\maketitle\subsection{\label{sec23} $m_t$-spectra}

\begin{figure}
\includegraphics[width=1.0\textwidth]{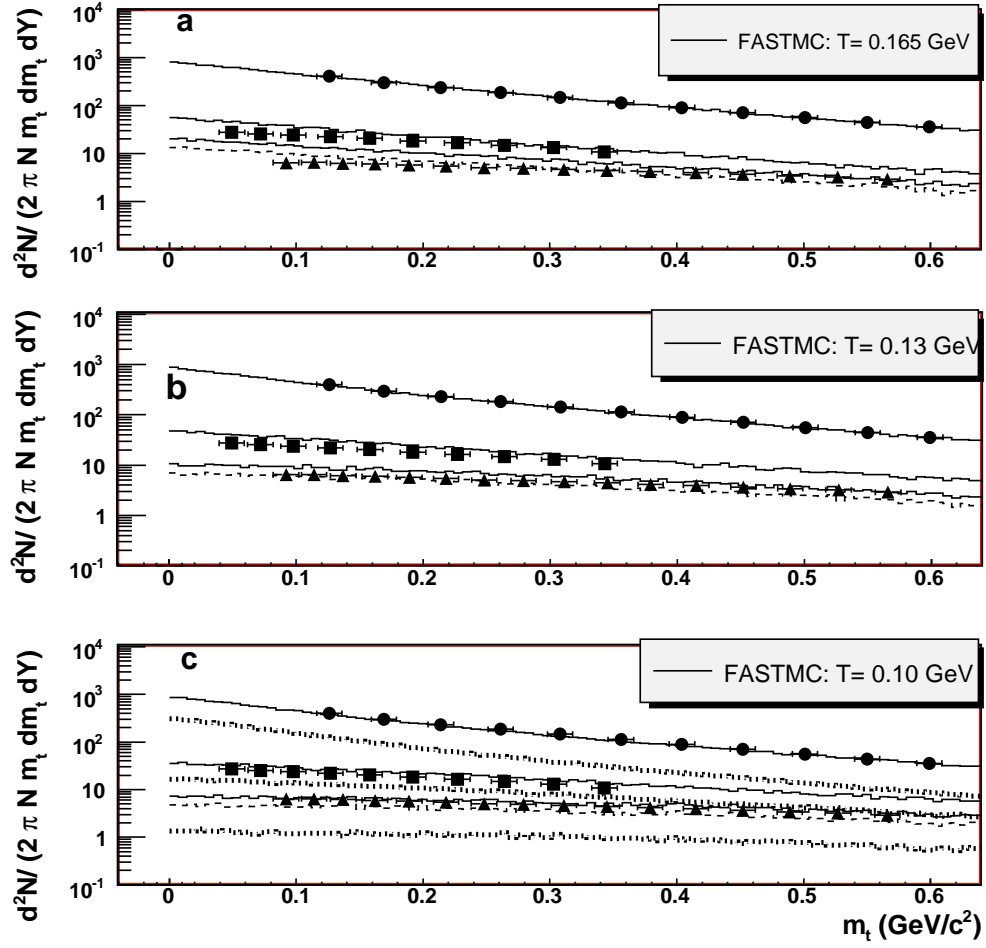}
\caption{
$m_t$-spectra (in $c^4$/GeV$^2$) measured by the STAR
Collaboration \cite{STAR_mt} for
$\pi^{+}$ (circles), $K^{+}$ (squares) and $p$  (up-triangles)at $0-5 \%$ centrality   in
comparison with the model calculations
at $T^{\rm th}=0.165 (a), 0.130 (b), 0.100 (c)$~GeV ,
with the parameters from Table \ref{tab:table1},
for protons weak decays are taken into account (solid lines);
for protons weak decays are not taken into account
 (dashed lines).
The direct
 $\pi^{+}$ , $K^{+}$ and $p$ contributions
are shown on {\bf(c)} by dotted lines.
\label{fig:mt_compar}
}
\end{figure}

In Fig.~\ref{fig:mt_compar}  the $m_t$-spectra
measured by the STAR
Collaboration~\cite{STAR_mt} at $0-5 \%$ centrality
are shown for $\pi^{+}$, $K^{+}$ and $p$
in comparison with the model calculations
under the assumption of
the common chemical and thermal freeze-out
at $T^{\rm th}=T^{\rm ch}=0.165$~GeV (Fig.~\ref{fig:mt_compar}(a))
and under the assumption that the thermal
freeze-out at $T^{\rm th}=0.100,0.130$~GeV
occurs after the chemical one (Fig.~\ref{fig:mt_compar}(b, c)).

The correction on weak decays was introduced by the STAR Collaboration
in pion spectra only \cite{STAR_mt}.
It was approximately $12 \%$ and was estimated from the measured $K_{s}^{0}$ and $\Lambda$
decays. In Ref.~\cite{STAR_mt} the STAR Collaboration doesn't introduce the
weak decay correction in proton spectra.
To reproduce the STAR weak decay correction procedure, we excluded pions from $K_{s}^{0}$ and $\Lambda$
decays from pions $m_t$-spectra in Fig.~\ref{fig:mt_compar}.
The contribution of weak decays in the simulated proton spectra
can be estimated from Fig.~\ref{fig:mt_compar}
by comparison of the solid lines (protons from $K_{s}^{0}$ and $\Lambda$ decays are included)
and the dashed lines (without contribution of protons from the weak decays).
The model parameters at different temperatures are presented in Table~\ref{tab:table1}.
The parameters were optimized this way to obtain the good description
of the pion $m_t$-spectra and the correlation radii.
The best description of the $m_t$-spectra was achieved at $T^{\rm th}=0.100$~GeV
(Fig.~\ref{fig:mt_compar}(c)).

\begin{figure}
\includegraphics[width=1.0\textwidth]{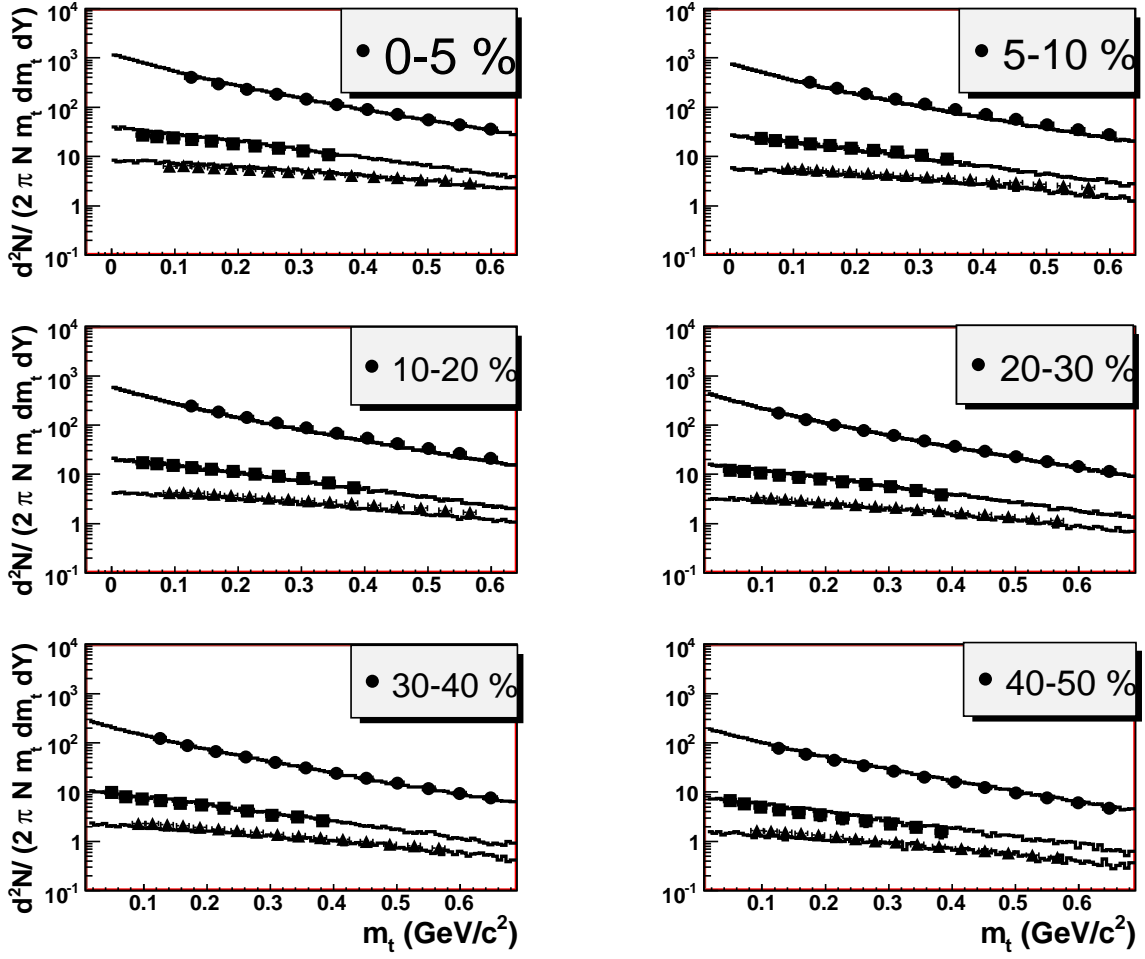}
\caption{
$m_t$-spectra (in $c^4$/GeV$^2$) measured by the STAR
Collaboration  \cite{STAR_mt} for
$\pi^{+}$ (circles), $K^{+}$ (squares) and $p$ (up-triangles) at different centralities in
comparison with our fast MC calculations at $T^{\rm th}=0.100$~GeV (solid lines)
with the parameters from Table \ref{tab:table1} and Table \ref{tab:table2}.
\label{fig:mt_all}
}
\end{figure}

The same set of parameters ${T, \rho_{u}^{\max}, R}$ and $\tau$ which was
determined for central collisions (Table \ref{tab:table1})
was used for noncentral ones.
The additional parameters of the model for noncentral collisions were
coordinate and momentum asymmetries: $\epsilon$ and $\delta$ (Table \ref{tab:table2}).
At the freeze-out moment we consider them as free parameters
because we do not trace the evolution here.
The influence of the choice of $\epsilon$ and $\delta$ on $m_t$-spectra averaged over azimuthal angle $\varphi$
is negligible.
The decrease of the effective volume in noncentral collisions (Eq.~\ref{HD3})
due to nonzero values of $\epsilon$ and $\delta$
allows us to obtain the correct absolute normalization of $m_t$-spectra without
introduction of the additional parameters.
In Fig.~\ref{fig:mt_all}  the $m_t$-spectra
measured by the STAR
Collaboration~\cite{STAR_mt} are shown for $\pi^{+}$, $K^{+}$ and $p$ at centralities:
$0-5 \%, 5-10 \%, 10-20 \%, 20-30 \%, 30-40 \%, 40-50 \%$
in comparison with the model calculations which assume that the thermal freeze-out at $T^{\rm th}=0.1$~GeV
occurs after the chemical one (solid lines).
It appears that the procedure described in section~\ref{sec1}
allows one to achieve the absolute normalization of pion spectra correct
within $\sim 13 \%$.

\maketitle\subsection{\label{sec24} Elliptic flow}

Following a standard procedure~\cite{voloshin96,
poskanzer98} we make a Fourier expansion of the hadron distribution in the
azimuthal angle $\varphi$ at mid-rapidity:
\begin{equation}
\label{dN}
\frac{dN}{d^{2}p_{t}dy}= \frac{dN}{2\pi p_t dp_t dy} (1+2v_{2}\cos2\varphi
+2v_{4}\cos4\varphi+...).
\end{equation}

The elliptic flow coefficient, $v_2$, is defined as the
second order Fourier coefficient,
\begin{equation}
\label{v2}
v_{2}= \frac{
\int_{0}^{2 \pi} d \varphi \cos 2(\varphi-\psi_{R})\frac{d^3N}{dy d\varphi p_t dp_t}}
{\int_{0}^{2 \pi} d \varphi \frac{d^3N}{dy d\varphi p_t dp_t}},
\end{equation}
where $\psi_{R}$ is the reaction plane angle (in our generation $\psi_{R}=0$), $y$
and $p_t$ are the rapidity and transverse momentum of particle under
consideration, respectively.

\begin{figure}
\includegraphics[width=1.0\textwidth]{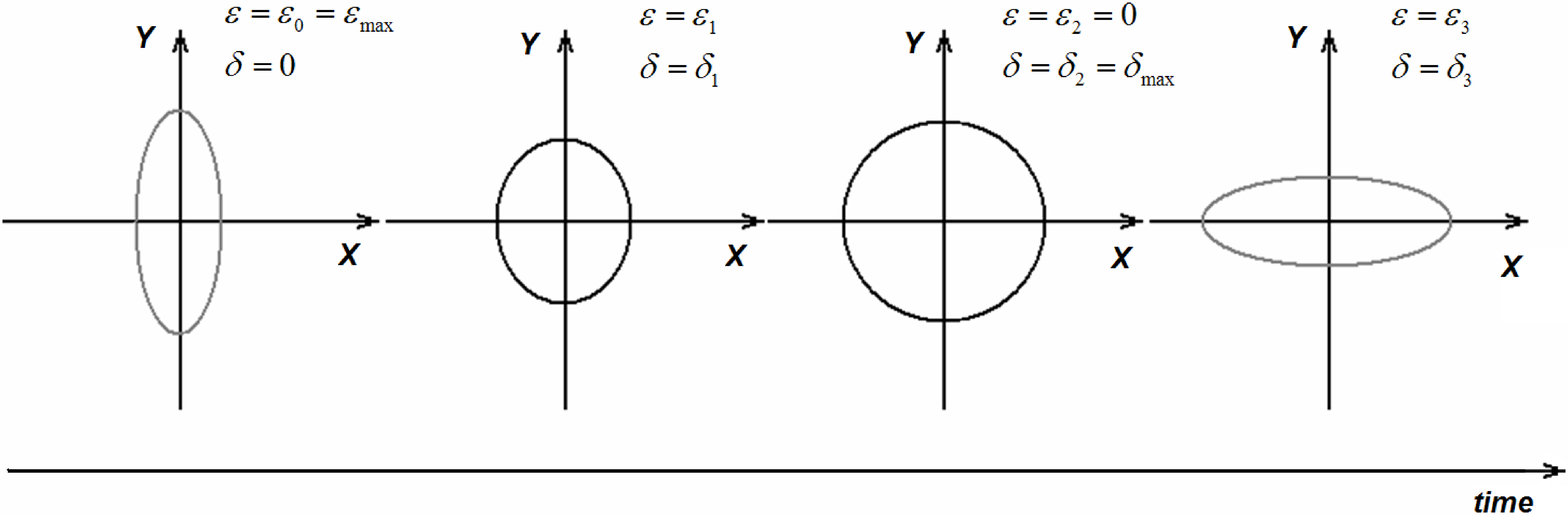}
\caption{
 The typical hydrodynamic evolution scenario.
\label{fig:hydro_evolution}
}
\end{figure}

The value of $v_2$ is an important signature of the physics occurring in
heavy ion collisions.  According to the typical hydrodynamic scenario
shown in Fig.~\ref{fig:hydro_evolution},
the elliptic flow is generated mainly during the high density phase of the
fireball evolution. The system
driven by the internal pressure gradients expands more strongly
in its short direction (into the direction of the impact
parameter
$x$ in Fig.~\ref{fig:hydro_evolution}, which is chosen as a ``positive''
direction)
than in the perpendicular one (``negative'' direction, $y$ in Fig.~\ref{fig:hydro_evolution}) where the pressure gradients are smaller.
Figure~\ref{fig:hydro_evolution} illustrates qualitatively that the initial spacial anisotropy of the system
disappears during the evolution, while the momentum anisotropy grows.
The developing of strong flow observed at RHIC
requires a short time scale and large pressure gradients, which are
characteristics of a strongly interacting system. The reason for the
generation of $v_2$ at the early times is that the system should be hot and dense,
when the system cools and become less
dense the developing of the large pressure gradients becomes impossible.
The elliptic flow
coefficient, $v_2$, depends on
the transverse momentum $p_t$, the impact parameter $b$ or centrality,
as well as, the type of the considered particle.
All these dependencies have
been measured at RHIC \cite{STAR_v2}.

\begin{figure}
\includegraphics[width=1.0\textwidth]{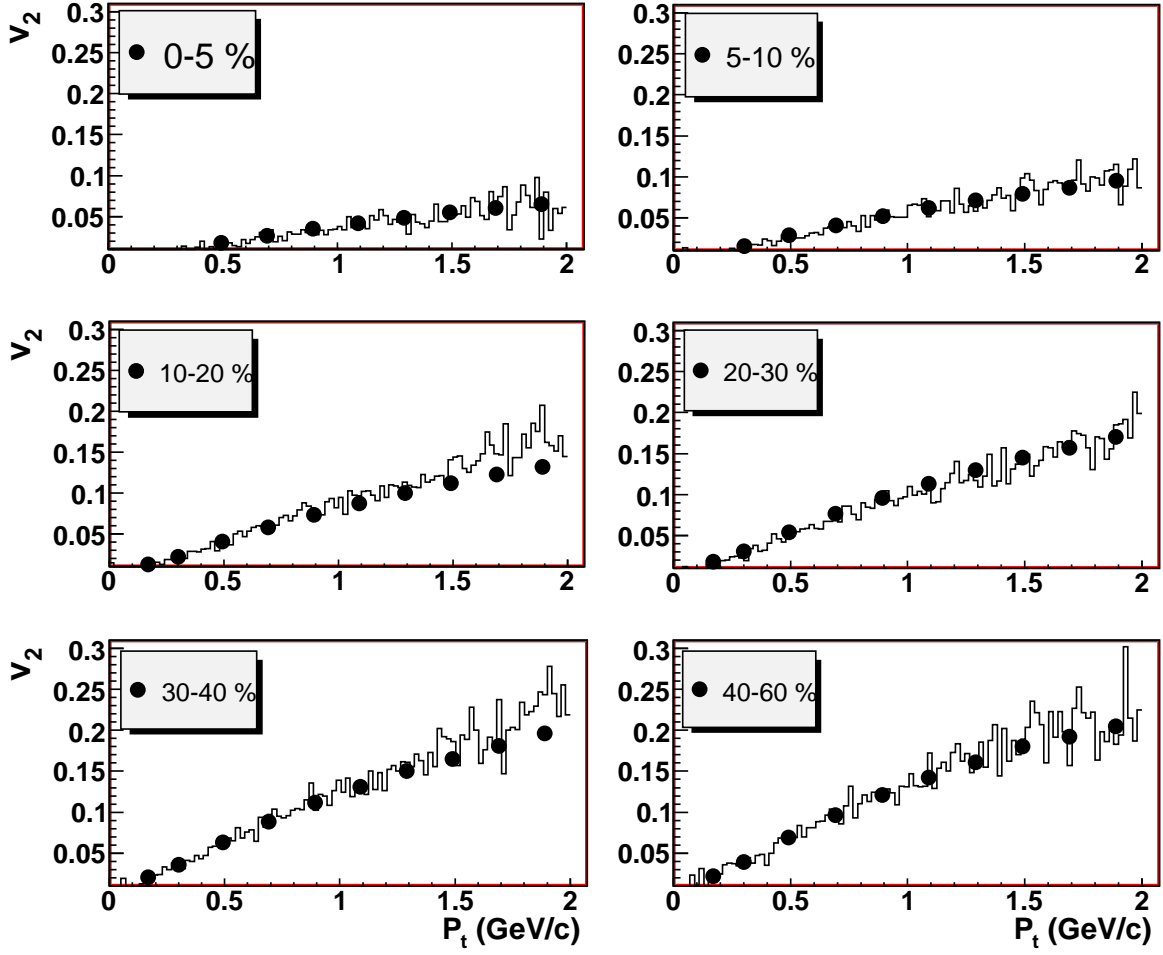}
\caption{
The $p_t$-dependence of $v_2$ measured by the STAR
Collaboration \cite{STAR_v2} (points) for charged particles at different centralities in comparison
with our fast MC calculations at $T^{\rm th}=0.100$~GeV (solid line)
with the parameters from Table \ref{tab:table1} and Table \ref{tab:table2}.
\label{fig:v2_all}
}
\end{figure}

The $p_t$-dependence of $v_2$ measured by the STAR
Collaboration~\cite{STAR_v2} for charged particles at centralities:
$0-5 \%, 5-10 \%, 10-20 \%, 20-30 \%, 30-40 \%, 40-60 \%$
is shown in Fig.~\ref{fig:v2_all}
in comparison with our MC calculations obtained
with the optimal model parameters from Table~\ref{tab:table2}.
The calculations were performed
under the assumption that thermal freeze-out at $T^{\rm th}=0.1$~GeV
occurs after the chemical one at $T^{\rm th}=0.165$~GeV.

\begin{figure}
\includegraphics[width=1.0\textwidth]{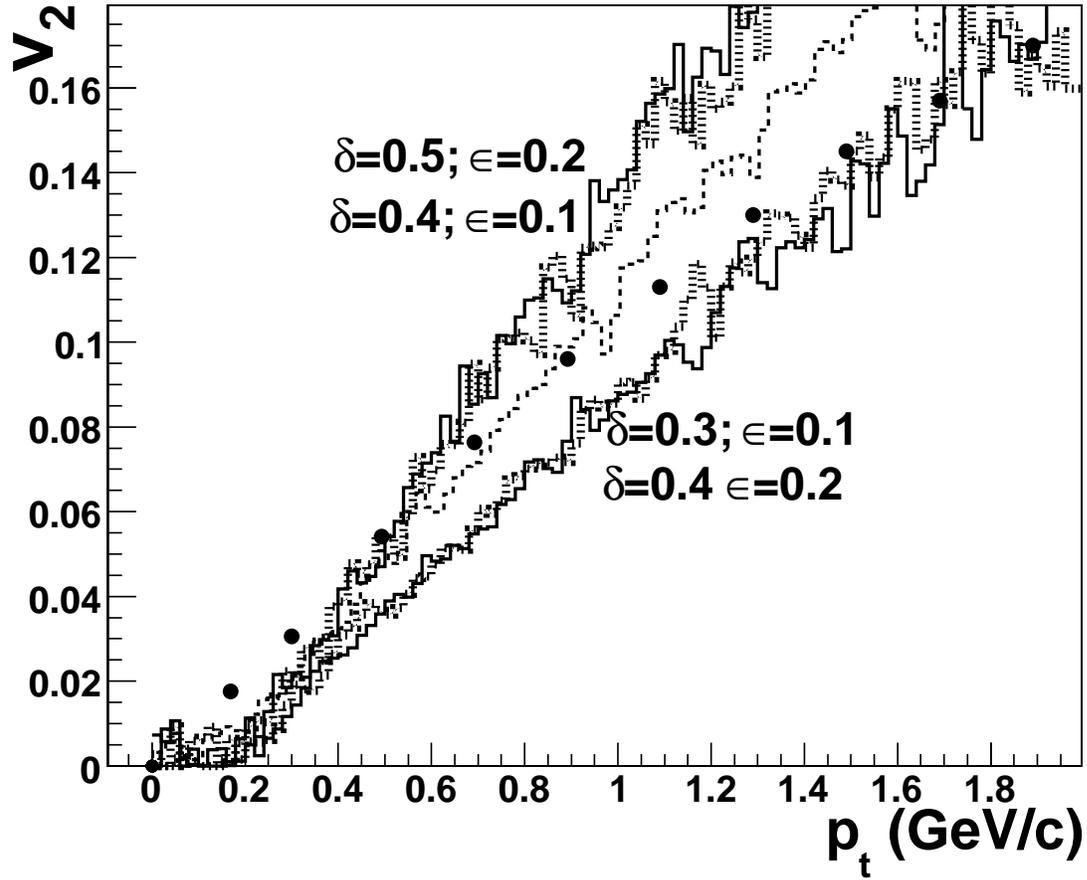}
\caption{
The $p_t$-dependence of $v_2$ measured by the STAR
Collaboration \cite{STAR_v2} (points) for charged particles
at centrality $20-30 \%$ in comparison with
our fast MC calculations under assumption of the
single freeze-out at $T^{\rm th}=T^{\rm ch}=0.165$~GeV.
The different sets of coordinate and momentum assymetries parameters were tried:
$\epsilon =0.1$, $\delta =0.3$ (solid line),
$\epsilon =0.2$, $\delta =0.4$ (dotted line),
$\epsilon =0.1$, $\delta =0.4$ (solid line),
$\epsilon =0.2$, $\delta =0.5$ (dotted line),
$\epsilon =0.15$, $\delta =0.4$ (dashed line)
\label{fig:v2_1}
}
\end{figure}

The calculations under the assumption of
the common chemical and
thermal freeze-out at $T^{\rm th}=T^{\rm ch}=0.165$~GeV
demonstrate not so good agreement with the experimental data at small
$p_t < 0.4$ GeV/$c$
for the centralities larger than $20 \% $;
irrespective of the choice of $\epsilon$ and $\delta$ one
cannot get a satisfactory description in the whole $p_t$-range (see e.g. Fig.~\ref{fig:v2_1}).

\maketitle\subsection{\label{sec25}
Correlation radii}

\begin{figure}
\includegraphics[width=1.0\textwidth]{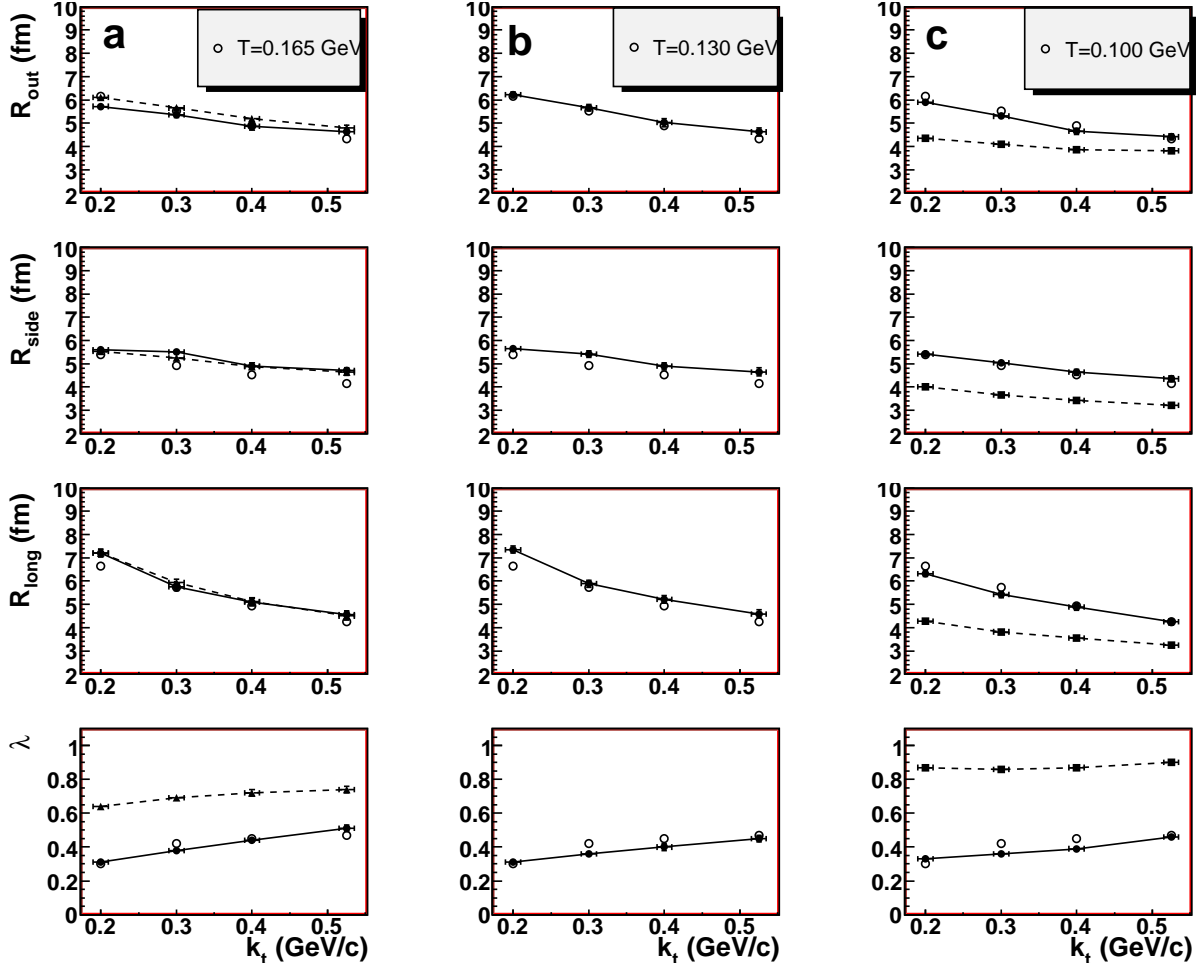}
\caption{
The $\pi^{+}$ correlation radii
at mid-rapidity in central Au+Au
collisions at $\sqrt s_{NN} = 200$~GeV from the STAR experiment
\cite{CF_STAR} (open circles) and  MC calculations
within the Bjorken-like model with the parameters presented in
Table~\ref{tab:table1}
in different intervals of the pair
transverse momentum $k_t$. The full calculation
with resonances (a), (b).
(a) single freeze-out $T^{\rm ch } =T^{\rm th}=0.165$~GeV,
 no weak decays (dashed
line), with weak decays (solid
line); (b) thermal
freeze-out at $T^{\rm th}=0.130$~GeV
occurs after the chemical one,
weak decays are taken into account (solid line);
(c) the full calculation with resonances,
weak decays are taken into account at $T^{\rm th}=0.100$~GeV
 (solid line), the direct pions only (dotted lines).
\label{fig:CFcentr_th}
}
\end{figure}

The parameters of the model presented in
Table~\ref{tab:table1} were optimized to
obtain the best description of the pion $m_t$-spectra and
the correlation radii in the following cases:
under the assumption of
the common chemical and thermal freeze-out
at $T^{\rm th}=T^{\rm ch}=0.165$~GeV
and under the assumption that the thermal
freeze-out at $T^{\rm th}=0.100,0.130$~GeV
occurs after the chemical one.
In Fig.~\ref{fig:CFcentr_th} the fitted correlation radii
$R_{\rm out}, R_{\rm side}$ and $R_{\rm long}$
are compared with those measured by the STAR Collaboration~\cite{CF_STAR}.
The three-dimensional correlation function was fitted with the standard Gaussian formula:
\begin{equation}
CF(p_{1},p_{2})= 1+\lambda\exp(-R_\mathrm{out}^2q_\mathrm{out}^2
-R_\mathrm{side}^2q_\mathrm{side}^2
-R_\mathrm{long}^2q_\mathrm{long}^2), \label{cf3}
\end{equation}
where $\vec{q}=\vec{p_1}-\vec{p_2}=(q_{\rm out}, q_{\rm side}, q_{\rm long})$ is
the relative three-momentum of two identical particles with four-momenta $p_1$
and $p_2$.
The form of Eq.~(\ref{cf3}) assumes azimuthal symmetry of
the production process \cite{pod83}. Generally, e.g., in the  case
of the correlation analysis with respect to the reaction plane, all
three cross terms $q_iq_j$ can be significant~\cite{Wiedemann_Heinz02}.
We will consider this case below.
We choose the longitudinal co-moving system
(LCMS) as the reference frame~\cite{cso91}. In  LCMS each pair is emitted
transverse to the
reaction axis so that the pair rapidity vanishes. The parameter
$\lambda$ measures the correlation strength.

The regime with the large temperature $T^{\rm th}=T^{\rm ch}=0.165$~GeV
was tested in Ref.~\cite{FASTMC}. We have repeated this test here with
the new resonances table
and the additional parameter $\Delta \tau$ (Fig.~\ref{fig:CFcentr_th}(a), dashed line).
We have found that these modifications lead to a better
description of the correlation radii.
In Fig.~\ref{fig:CFcentr_th}(a, bottom) (dashed line) the intercept $\lambda$ is larger
than the experimental one, but
taking into account the secondary pions from the weak decays
essentially improves the description of the $\lambda$ (Fig.~\ref{fig:CFcentr_th}(a, bottom),
solid line).

In Fig.~\ref{fig:CFcentr_th}(b, c) we consider the lower
thermal freeze-out temperatures: $0.130, 0.100$~GeV.
The secondary pions coming from the weak decays were taken into account.

It is worth to note a good description of the
correlation radii (within $\sim 10 \%$ accuracy) altogether with
the absolute value of the $m_t$ spectra in the scenario with a low
temperature thermal freeze-out of chemically frozen
hadron-resonance gas. There are three important reasons for this
success. First, a relatively small (compared with dynamic models) effective volume of
the system $\sim \tau R^2$ that reduces the correlation radii.
Second, relatively large transverse flow in the model that further
reduces the radii. Third, rather large effective pion chemical
potential which is needed to describe the absolute value of the
pion spectra at relatively small effective volumes; it reduces
correlation radii at small $p_t$ and so makes their $m_t$
behavior flatter. This reduction happens due to vanishing of the
homogeneity length of Bose-Einstein distribution
for low-$p_t$ pions when the pion chemical
potential approaches the pion mass
(see also Ref.~\cite{pointBEC} for the analysis of the reduction of the
pion correlation radii near the
point of the Bose-Einstein condensation in static systems).
We do not consider here the question whether such conditions
could be realized in realistic dynamical models.

It should be noted that the description of the $k_t$-dependence
of the correlation radii has been
achieved within $\sim 10 \%$ accuracy
for all three considered thermal temperatures:
$T^{\rm th} = 0.165, 0.130, 0.100$~GeV.
However, at lower temperatures
there is more flexibility in the simultaneous description of
particle spectra and correlations
because the effective volume
isn't strictly fixed as it is in the case of the single freeeze-out
($T^{\rm th}=T^{\rm ch}=0.165$ GeV).
In present work,
we have not attempted to fit the model parameters
($T^{\rm th}$, $R$, $\tau$, $\mu_{\pi}^{\rm eff~th}$) since it
is rather complicated task requiring a lot of computer time.
We have performed only example calculations with several
sets of the parameters.

In noncentral collisions the measurement of azimuthally sensitive correlation radii
provides the additional information about the source shape.
For the corresponding femtoscopy formalism with respect to the reaction plane
see, e.g., \cite{wiedemann98, Wiedemann_Heinz02}.
In the absence of azimuthal symmetry, the three additional
cross terms contribute to the Gaussian parametrization
of the correlation function in
Eq.~(\ref{cf3}):
\begin{equation}
CF(p_{1},p_{2})= 1+\lambda\exp(-R_\mathrm{o}^2q_\mathrm{out}^2
-R_\mathrm{s}^2q_\mathrm{side}^2
-R_\mathrm{l}^2q_\mathrm{long}^2
-2R_\mathrm{os}^2q_\mathrm{out}q_\mathrm{side}
-2R_\mathrm{ol}^2q_\mathrm{out}q_\mathrm{long}
-2R_\mathrm{sl}^2q_\mathrm{side}q_\mathrm{long}). \label{cf3ad}
\end{equation}
In the boost-invariant case, the transverse-longitudinal
cross terms $R_{\rm ol}^2$ and $R_{\rm sl}^2$ vanish in the LCMS frame, while
the important out-side $R_{\rm os}^2$ cross term is present.

In the Gaussian approximation,
the radii in the Eq.~(\ref{cf3ad}) are related to space-time variances
via the set of equations \cite{wiedemann98, Wiedemann_Heinz02}:
\begin{equation}
\begin{array}{c}
R_\mathrm{s}^2 =
1/2(\langle \widetilde{x}^2 \rangle+\langle \widetilde{y}^2\rangle)
-1/2(\langle \widetilde{x}^2 \rangle-\langle \widetilde{y}^2 \rangle) \cos(2\Phi)-
\langle \widetilde{x}\widetilde{y} \rangle\sin(2\Phi),
\\
R_\mathrm{o}^2 =
1/2(\langle \widetilde{x}^2 \rangle+\langle \widetilde{y}^2\rangle)
+1/2(\langle \widetilde{x}^2 \rangle-\langle \widetilde{y}^2 \rangle) \cos(2\Phi)
+\langle \widetilde{x}\widetilde{y} \rangle\sin(2\Phi))
\\
-2\beta_{\perp} (\langle \widetilde{t}\widetilde{x} \rangle\cos(\Phi)+
\langle \widetilde{t}\widetilde{y} \rangle\sin(\Phi)) + \beta_{\perp}^{2} \langle
\widetilde{t}^2\rangle,
\\
R_\mathrm{l}^2 = \langle \widetilde{z}^2 \rangle -2 \beta_l \langle \widetilde{t}\widetilde{z} \rangle
+ \beta_{l}^{2} \langle \widetilde{t}^2\rangle, \\
R_\mathrm{os}^2 =
\langle \widetilde{x}\widetilde{y} \rangle\cos(2\Phi)
-1/2(\langle \widetilde{x}^2 \rangle-\langle \widetilde{y}^2 \rangle) \sin(2\Phi)
+\beta_{\perp} (\langle \widetilde{t}\widetilde{x} \rangle\sin(\Phi)-
\langle \widetilde{t}\widetilde{y} \rangle\cos(\Phi)),

\end{array}
\end{equation}
where $\beta_l=k_z/k^0$, $\beta_{\perp}=k_{\perp}/k^0$ and
$\Phi=\angle (\vec{k_{\perp}},\vec{b})$
is the azimuthal angle of the pair three-momentum $\vec{k}$ with respect to
the reaction plane $z$-$x$ determined by the longitudinal direction and the
direction of the impact parameter vector
 $\vec{b}=(x,0,0)$;
the space-time coordinates $\widetilde{x}^{\mu}$
are defined relative to the effective source
center $\langle x^{\mu}\rangle$:
 $\widetilde{x}^{\mu}=x^{\mu}-\langle x^{\mu}\rangle $.
The averages are taken with the source emission function
$S(t, \vec{x}, k)$,
\cite{wiedemann98}:
\begin{equation}
\langle f(t, \vec{x}) \rangle =\frac{\int{d^4 x f(t, \vec{x})S(t, \vec{x},k)}}
{\int{d^4 x S(t, \vec{x},k)}}.
\end{equation}

\begin{figure}
\includegraphics[width=1.0\textwidth]{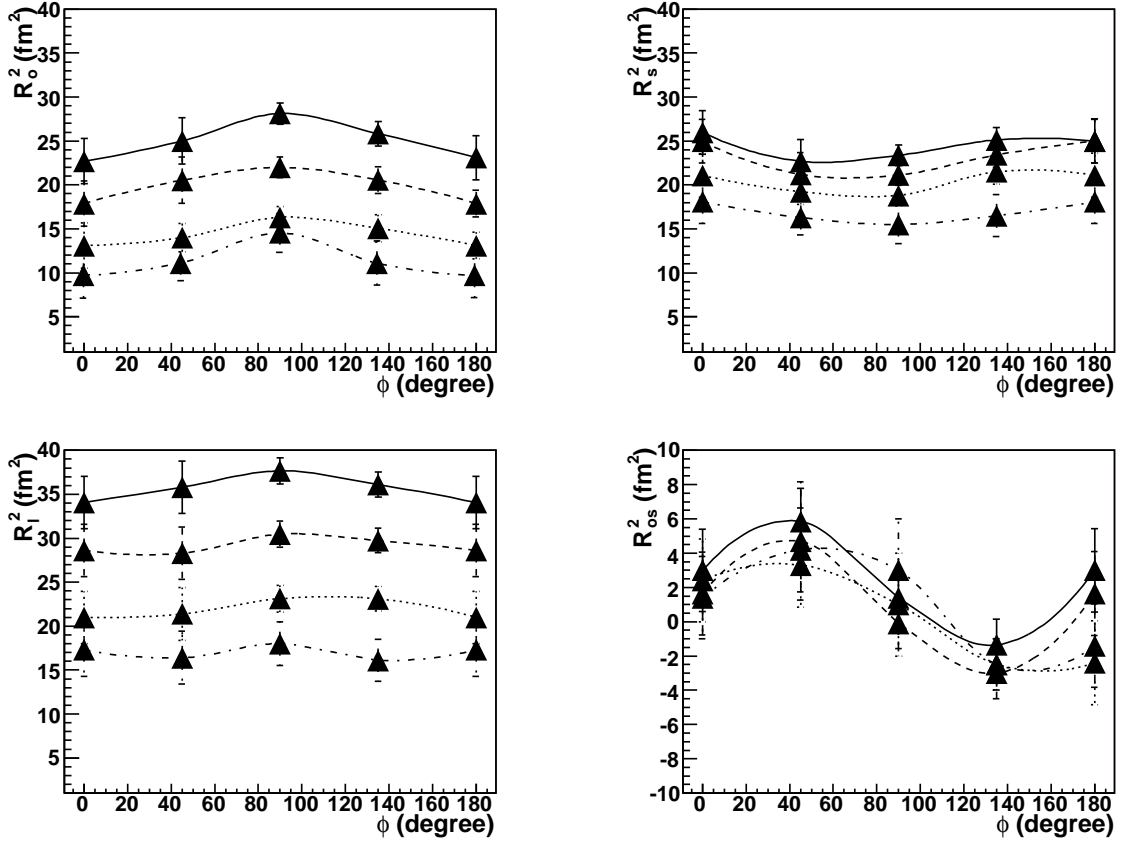}
\caption{
Simulated with FASTMC squared correlation radii versus
the azimuthal angle $\Phi$ of the $\pi^{+} \pi^{+}$ pair with respect to
the reaction plane,
20-30 $\%$ centrality events in $k_{T}$ (GeV/$c$) intervals:
$0.15<k_{T}<0.25$ (solid line), $0.25<k_{T}<0.35$ (dashed line),
$0.35<k_{T}<0.45$ (dotted line), $0.45<k_{T}<0.60$ (dotted-dashed line).
simulation was done with the special set of parameters:
$T^{\rm th}=0.1$~GeV, $\rho_{u}^{max}(b=0)=1.0$;
$R(b=0)=$11.5~fm, $\tau=$7.5~fm/$c$, $\Delta \tau=$0.~fm/$c$,
$\epsilon=0.1$ and $\delta=0.25$,
weak decays were not taken into account.
\label{fig:asHBT}
}
\end{figure}

The illustrative calculations of the correlation radii
as a function of the azimuthal angle $\Phi$
were done with the following fast MC parameters:
$T^{\rm th}=0.1$~GeV, $\rho_{u}^{max}(b=0)=1.0$;
$R(b=0)=$11.5~fm, $\tau=$7.5~fm/$c$, $\Delta \tau=$0.~fm/$c$,
$\epsilon=0.1$ and $\delta=0.25$.
The azimuthal dependence of the correlation radii in
different $k_t$ intervals is shown in Fig.~\ref{fig:asHBT}.

The $R_{s}^{2}$
oscillates downward, in the same phase as "RHIC" source extended
out of plane \cite{STAR-asCFs}, which means the larger sideward radius viewed from
the $x$-direction (in the reaction plane),
than from $y$-direction (out-of plane).
The source has small coordinate asymmetry $\epsilon=0.1$, it is almost round
(as in Fig.~\ref{fig:hydro_evolution} step 3), however
the emission zone, or ``homogeneity
region'', varies with $\Phi$ because of
the non-isotropic flow.

\maketitle
\section{\label{sec3} Conclusions }
We have developed a MC simulation procedure and the corresponding
C++ code, that allows a fast
realistic description of multiple hadron production both in
central and noncentral relativistic heavy ion collisions.
A high generation speed and an easy control through input
parameters make our MC generator code particularly useful
for detector studies.
As options, we have implemented two freeze-out scenarios with
coinciding and with different chemical and thermal freeze-outs.
We have compared the RHIC experimental data with our MC
generation results obtained within the single and separated freeze-out scenarios with
Bjorken-like freeze-out surface parameterization.

Fixing the temperatures of the chemical and thermal freeze-out at
0.165 GeV and 0.100 GeV respectively, and, using the same set of the model parameters
as for the central collisions, we have described
single particle spectra at different centralities with the absolute normalization
correct
within $\sim 13 \%$.

The comparison of the RHIC $v_2$ measurements with our MC
generation results shows that the scenario with two separated freeze-outs
is more favorable for the description of the
$p_t$-dependence of the elliptic flow.

The description of the $k_t$-dependence of the correlation radii
has been
achieved within $\sim 10 \%$ accuracy. The experimentally observed
values of the correlation strength parameter $\lambda$ has been reproduced
due to the account of the weak decays.

The analysis of the azimutal
dependence of the correlation radii indicates that
the
source considered in the model
oscillates downward, in the same phase as "RHIC" source extended
out of plane.

The achieved understanding of the reasons leading to a good
simultaneous description of particle spectra, elliptic flow
and femtoscopic correlations within
the considered simple model could be useful for building of the complete dynamic
picture of the matter evolution in A+A collisions.

\begin{acknowledgments}
We would like to thanks
A.~Kisiel, W.~Broniowski and W.~Florkowski
for the permission to use the particle data table from SHARE
used in their THERMINATOR code.
We would like to thank Eu.~Zabrodin, B.V.~Batyunia and L.I.~Sarycheva
for useful discussions.
The research has been carried out within the scope of the ERG
(GDRE): Heavy ions at ultra-relativistic energies - a European
Research Group comprising IN2P3/CNRS, Ecole des Mines de Nantes,
Universite de Nantes, Warsaw University of Technology, JINR Dubna,
ITEP Moscow and Bogolyubov Institute for Theoretical Physics NAS of
Ukraine.
The investigations have been partially supported by the
IRP AVOZ10480505, by the Grant Agency of the Czech
Republic under Contract No. 202/07/0079 and by the grant LC07048 of the
Ministry of Education of the Czech Republic
and by
Award No. UKP1-2613-KV-04 of the U.S. Civilian
Research and Development Foundation
(CRDF) and Fundamental Research State Fund of Ukraine, Agreement No.
F7/209-2004
and also by Award of Physics and Astronomics Division of NASU /017U000396
(19.12.06), Fundamental Research State Fund of Ukraine, Agreement No.
F25/718-2007
and Bilateral award DLR(Germany)-MESU (Ukraine) for UKR 06/008 Project, 
and by grant N 08-02-91001-CERN-a of Russian Foundation for Basic Research.

\end{acknowledgments}

\clearpage

\end{document}